\documentclass{article} % Changed from svjour3 to article
\newcommand{\keywords}[1]{\textbf{Keywords:} #1}

\usepackage[utf8]{inputenc}
\usepackage{amsmath,amssymb,amsfonts}
\usepackage{amsthm}
\newtheorem{theorem}{Theorem}
\usepackage{bm}
\usepackage{hyperref}

\usepackage{authblk}
\newcommand{\email}[1]{\texttt{#1}}

\title{Bell, Spinors, and the Impossibility of a Classical Spin-Vector Model}

\author{G.~A.~Koroteev \\%
   \email{greg.koroteev@gmail.com}}

\begin{document}
\maketitle

\begin{abstract}
We revisit the Bell--CHSH scenario for two spin-\(\tfrac{1}{2}\) particles and isolate the precise algebraic origin of the Bell contradiction. On the quantum side, spin-\(\tfrac{1}{2}\) is described by a noncommutative spinor (Clifford) algebra acting on the Hilbert space of two spin-\(\tfrac{1}{2}\) particles, with the singlet state yielding the usual correlation \(E(a,b) = -\,a\cdot b\) and Tsirelson's bound \(2\sqrt{2}\). On the classical side, the standard Bell assumptions amount to describing all measurement outcomes as \(\{\pm1\}\)-valued random variables on a single Kolmogorov probability space, i.e.\ elements of a commutative algebra \(\mathcal{C}(\Lambda)\).

We show that there is no representation of the spinor algebra of spin-\(\tfrac{1}{2}\) (with its singlet state and locality structure) into any such commutative Kolmogorov algebra that preserves the \(\{\pm1\}\) spectra of local spin components and the singlet correlations entering the CHSH expression, under the standard Bell assumptions of locality (factorization) and measurement independence. In this sense, the Bell--CHSH contradiction is exhibited as an algebraic mismatch between a noncommutative spinor/Clifford description of spin and the classical assumption of a single global Kolmogorov space supporting all outcomes. In the language of quantum probability, this is a C\(^*\)-algebraic reformulation of the known fact that the singlet correlations admit no local hidden-variable model with jointly distributed outcomes on one probability space.

We also give an explicit realization of the same spinor structure within the author's Quantum Index Algebra (QIA) framework, where locality appears as disjoint index slots and the singlet state as a simple index cocycle.

\end{abstract}

\section{Introduction}
\label{sec:intro}

Bell's theorem and its CHSH form \cite{Bell1964,CHSH1969} show that no local hidden-variable theory can reproduce all predictions of quantum mechanics for entangled systems. For two spin-\(\tfrac{1}{2}\) particles in the singlet state, quantum mechanics predicts a correlation
\begin{equation}
  E(a,b) = -\,a\cdot b,
\end{equation}
and a maximal violation of the CHSH inequality
\begin{equation}
  |S| \le 2
\end{equation}
up to Tsirelson's bound \(2\sqrt{2}\) \cite{Tsirelson1980}. Loophole-free experiments with photons, ions and solid-state systems have confirmed this violation to high statistical confidence \cite{Aspect1982,Hensen2015,Giustina2015,Shalm2015}. This is often summarized by the slogan that ``nature is nonlocal'' or that ``realism and locality cannot both hold'', under the usual assumptions of measurement independence and the absence of retrocausal or superdeterministic effects.

In the standard textbook presentation, the structure of the hidden-variable model is usually phrased in probabilistic and philosophical terms: realism (pre-existing values), locality (no faster-than-light influences), a single Kolmogorov probability space, and measurement independence. What is rarely isolated and emphasized is the \emph{algebraic} aspect of the classical model: all measurement outcomes for all settings are represented as random variables on one probability space \((\Lambda,\rho)\), hence as elements of a single commutative C\(^*\)-algebra \(\mathcal{C}(\Lambda)\). In particular, all spin components for all measurement directions are assumed to have simultaneous (though unmeasurable) pre-existing values.

On the quantum side, however, spin-\(\tfrac{1}{2}\) is not described by classical vectors but by spinors, acted upon by noncommuting operators that generate a Clifford (or Pauli) algebra. For a single particle, spin along a direction \(n\) is given by a matrix \(\Sigma(n)\) satisfying
\begin{equation}
  \Sigma(n)^2 = I,\qquad
  \Sigma(n)\Sigma(m) + \Sigma(m)\Sigma(n) = 2(n\cdot m)\,I,
\end{equation}
and different components (e.g.\ \(\Sigma(a_0)\) and \(\Sigma(a_1)\) for noncollinear \(a_0,a_1\)) do not commute. For two particles, locality is encoded as a tensor-product structure \(H = S_A \otimes S_B\) with local operators
\begin{equation}
  \Sigma_A(n)=\Sigma(n)\otimes I, \qquad \Sigma_B(m)=I\otimes \Sigma(m),
\end{equation}
acting on disjoint factors. The singlet state \(|\Psi\rangle\) lives in this \emph{spinor} framework and yields the observed CHSH violation.

The central observation of this paper is that the Bell--CHSH contradiction for spin-\(\tfrac{1}{2}\) can be expressed entirely as an algebraic non-embeddability statement:
\begin{equation}
  \mathcal{A}_{\mathrm{spin}} \not\hookrightarrow \mathcal{C}(\Lambda).
\end{equation}
Here \(\mathcal{A}_{\mathrm{spin}}\) denotes the unital C\(^*\)-algebra generated by the four local Pauli operators \(\Sigma_A(a_0),\Sigma_A(a_1),\Sigma_B(b_0),\Sigma_B(b_1)\) (and the identity) acting on \(\mathbb{C}^2\otimes\mathbb{C}^2\); for generic directions this is the full matrix algebra \(M_4(\mathbb{C})\). The algebra \(\mathcal{C}(\Lambda)\) is any commutative C\(^*\)-algebra of random variables on a hidden-variable space \(\Lambda\). There is no unital \(^*\)-homomorphism \(\Phi:\mathcal{A}_{\mathrm{spin}}\to\mathcal{C}(\Lambda)\) that simultaneously
(i) represents each local spin component as a \(\{\pm1\}\)-valued random variable,
and (ii) preserves the four pairwise singlet correlations entering the CHSH expression,
under the standard Bell assumptions of locality (factorization) and measurement independence. In other words, the spinor/Clifford spin algebra with the singlet state cannot be implemented as a classical Kolmogorov model with jointly distributed outcomes.

From this point of view, the Bell--CHSH contradiction is not primarily about choosing between locality and realism in the abstract. It is about the mismatch between
(a) the noncommutative spinor/Clifford algebra that actually describes spin-\(\tfrac{1}{2}\), and
(b) the commutative Kolmogorov algebra implicitly assumed in the classical hidden-variable model. The contradiction arises when one tries to treat spin-\(\tfrac{1}{2}\) as if it were a family of commuting \(\{\pm1\}\)-valued random variables on a single probability space, instead of noncommuting spinor operators in a Clifford (or Quantum Index Algebra) module.

In the QIA framework \cite{QIA_ref}, spin operators are realized as combinations of elementary \(2\times 2\) VC-blocks acting on index codes, and entangled states such as the singlet correspond to simple index cocycles. Locality is encoded as strict separation of index slots; all nonclassical features are traced to the cocycle and the noncommutative spin algebra. This representation allows us to restate the Bell--CHSH scenario without additional interpretational assumptions, as a purely algebraic incompatibility between a QIA spin module and any classical model based on a single Kolmogorov probability space.

\subsection*{Scope and contribution}

The non-embeddability statement itself is mathematically equivalent to the well-known result that no single joint probability distribution for all four measurement outcomes can reproduce the singlet correlations \cite{Fine1982,Pitowsky1989,AbramskyBrandenburger2011}. Our contribution lies in exhibiting the obstruction as an explicit algebraic mismatch between the concrete spinor/Clifford structure of spin-\(\tfrac{1}{2}\) and the commutative random-variable picture, with the QIA framework providing an especially transparent index-based realization of the same structure.

The rest of the paper is organized as follows. In Sec.~\ref{sec:CHSH} we recall the standard Bell--CHSH setup in a minimal algebraic form. In Sec.~\ref{sec:main_theorem} we formulate the spinor algebra and state the main non-embeddability theorem. In Sec.~\ref{sec:spinor_algebra} we describe the spinor/Clifford and QIA realization of spin-\(\tfrac{1}{2}\). In Sec.~\ref{sec:kolmogorov} we analyze the role of Kolmogorov probability and global joint distributions. We conclude in Sec.~\ref{sec:conclusion} with a brief conceptual discussion.

\section{Standard Bell--CHSH Setup}
\label{sec:CHSH}

We briefly recall the CHSH scenario in a form that will be convenient for the algebraic analysis below.

Consider a source emitting pairs of spin-\(\tfrac{1}{2}\) particles toward two distant observers, Alice and Bob. In each run, Alice chooses one of two measurement settings \(a_0,a_1\) (unit vectors in \(\mathbb{R}^3\)), and Bob chooses one of two settings \(b_0,b_1\). Each measurement yields an outcome
\begin{equation}
  A(a_x) \in \{\pm 1\}, \qquad B(b_y) \in \{\pm 1\},
  \qquad x,y \in \{0,1\}.
\end{equation}
Correlations are defined as expectation values
\begin{equation}
  E(a_x,b_y) := \mathbb{E}[A(a_x) B(b_y)].
\end{equation}

On the classical side, the standard Bell assumptions are:
\begin{enumerate}
  \item \emph{Hidden variables (realism):} There exists a hidden variable \(\lambda \in \Lambda\) such that
    \begin{equation}
      A(a_x,\lambda) \in \{\pm1\}, \qquad B(b_y,\lambda) \in \{\pm1\},
    \end{equation}
    and the observed outcomes are determined (possibly stochastically) via a distribution \(\rho\) over \(\Lambda\).
  \item \emph{Locality (factorization):} Alice's outcome depends on \((a_x,\lambda)\) only, and Bob's on \((b_y,\lambda)\) only; in particular,
    \begin{equation}
      P(A,B|a_x,b_y,\lambda) = P(A|a_x,\lambda)\,P(B|b_y,\lambda).
    \end{equation}
    At the level of deterministic response functions, this means \(A(a_x,\lambda)\) is independent of \(b_y\), and \(B(b_y,\lambda)\) is independent of \(a_x\).
  \item \emph{Single Kolmogorov probability space:} All random variables \(A(a_0,\lambda),A(a_1,\lambda),B(b_0,\lambda),B(b_1,\lambda)\) live on one probability space \((\Lambda,\rho)\), so that
    \begin{equation}
      E_{\mathrm{cl}}(a_x,b_y) = \int_{\Lambda} A(a_x,\lambda)\,B(b_y,\lambda)\,\rho(d\lambda).
    \end{equation}
  \item \emph{Measurement independence (free choice):} The distribution \(\rho\) does not depend on which pair \((a_x,b_y)\) is chosen in a given run.
\end{enumerate}
These assumptions are typically summarized as ``local realism with standard Kolmogorov probability.''

Define, for each \(\lambda\), the classical CHSH combination
\begin{equation}
  S_{\mathrm{cl}}(\lambda)
  := A(a_0,\lambda) B(b_0,\lambda)
   + A(a_0,\lambda) B(b_1,\lambda)
   + A(a_1,\lambda) B(b_0,\lambda)
   - A(a_1,\lambda) B(b_1,\lambda).
\end{equation}
Since each factor is \(\pm1\), a simple enumeration shows that
\begin{equation}
  S_{\mathrm{cl}}(\lambda) = \pm 2 \qquad \Rightarrow \qquad |S_{\mathrm{cl}}(\lambda)| \le 2,
\end{equation}
for all \(\lambda\). Averaging over \(\rho\) yields the CHSH inequality
\begin{equation}
  |\langle S\rangle_{\mathrm{cl}}| := \left|\int_{\Lambda} S_{\mathrm{cl}}(\lambda)\,\rho(d\lambda)\right|\le 2.
\end{equation}
This bound follows purely from the existence of four \(\{\pm1\}\)-valued random variables on a single Kolmogorov space \((\Lambda,\rho)\), and it is independent of any details of the spin model.

On the quantum side, for two spin-\(\tfrac{1}{2}\) particles in the singlet state and with appropriate choices of \(a_0,a_1,b_0,b_1\), one finds a CHSH expectation
\begin{equation}
  |\langle S\rangle_{\mathrm{qm}}| = 2\sqrt{2},
\end{equation}
which violates the classical bound. This is the usual form of Bell's theorem: no local hidden-variable theory based on a single Kolmogorov space with factorization and measurement independence can reproduce the quantum predictions.

In the following sections we will:
(i) formulate the quantum side using the spinor/Clifford algebra of spin-\(\tfrac{1}{2}\),
(ii) interpret the classical side as a commutative C\(^*\)-algebra \(\mathcal{C}(\Lambda)\) of random variables,
and (iii) state and prove a non-embeddability theorem relating the two.

\section{Spinor Algebra of Spin and Main Non-Embeddability Result}
\label{sec:main_theorem}

In this section we formulate the algebraic structures on the quantum and classical sides and state the main non-embeddability theorem. Detailed constructions (spin as bivector in a Clifford algebra, QIA/VC-block representation) will be given in Sec.~\ref{sec:spinor_algebra}.

\subsection{Quantum spin algebra for two spin-\texorpdfstring{$\tfrac{1}{2}$}{1/2} particles}

Let \(S \cong \mathbb{C}^2\) be the spinor space for a single spin-\(\tfrac{1}{2}\) particle. For each unit vector \(n\in S^2\subset\mathbb{R}^3\), let \(\Sigma(n)\) be a \(2\times2\) matrix acting on \(S\) such that
\begin{equation}
  \Sigma(n)^2 = I, \qquad
  \Sigma(n)\Sigma(m) + \Sigma(m)\Sigma(n) = 2(n\cdot m)\,I,
\end{equation}
for all \(n,m\in S^2\). This realizes the usual Pauli/Clifford relations.

For two particles, the Hilbert space is
\begin{equation}
  H := S_A \otimes S_B,
\end{equation}
and we define local spin operators by
\begin{equation}
  \Sigma_A(n) := \Sigma(n)\otimes I,
  \qquad
  \Sigma_B(n) := I\otimes \Sigma(n).
\end{equation}
These satisfy
\begin{equation}
  [\Sigma_A(n),\Sigma_B(m)] = 0,
\end{equation}
for all \(n,m\), and each family \(\{\Sigma_A(n)\}\), \(\{\Sigma_B(n)\}\) separately obeys the single-particle Clifford relations.

Fix four measurement directions \(a_0,a_1,b_0,b_1\in S^2\). We define the \emph{spinor algebra of spin} for this Bell--CHSH configuration as the C\(^*\)-subalgebra
\begin{equation}
  \mathcal{A}_{\mathrm{spin}}
  := C^*\bigl(I,\, \Sigma_A(a_0),\Sigma_A(a_1),\Sigma_B(b_0),\Sigma_B(b_1)\bigr)
  \subset B(H),
\end{equation}
generated by these local spin operators and the identity. For generic choices of \(a_0,a_1,b_0,b_1\), this coincides with the full matrix algebra \(M_4(\mathbb{C})\).

Let \(|\Psi\rangle\in H\) be the singlet state characterized by
\begin{equation}
  \bigl(\Sigma_A(n) + \Sigma_B(n)\bigr)|\Psi\rangle = 0
  \qquad \forall n\in S^2.
\end{equation}
We define the \emph{spinor state} \(\omega_{\mathrm{spin}}\) on \(\mathcal{A}_{\mathrm{spin}}\) by
\begin{equation}
  \omega_{\mathrm{spin}}(T) := \langle\Psi, T\Psi\rangle.
\end{equation}
For each pair \((a_x,b_y)\) of directions we then have
\begin{equation}
  E_{\mathrm{spin}}(a_x,b_y)
  := \omega_{\mathrm{spin}}\bigl(\Sigma_A(a_x)\Sigma_B(b_y)\bigr)
  = -\,a_x\cdot b_y,
\end{equation}
and the corresponding CHSH operator
\begin{equation}
  S_{\mathrm{QIA}}
  := \Sigma_A(a_0)\Sigma_B(b_0)
   + \Sigma_A(a_0)\Sigma_B(b_1)
   + \Sigma_A(a_1)\Sigma_B(b_0)
   - \Sigma_A(a_1)\Sigma_B(b_1)
\end{equation}
has expectation
\begin{equation}
  \omega_{\mathrm{spin}}(S_{\mathrm{QIA}}) = 2\sqrt{2}
\end{equation}
for appropriate choices of the four directions.

\subsection{Classical Kolmogorov model}

On the classical side, let \((\Lambda,\rho)\) be a probability space. Classical observables are represented by bounded measurable functions on \(\Lambda\):
\begin{equation}
  \mathcal{C}(\Lambda) := L^\infty(\Lambda,\rho),
\end{equation}
with pointwise multiplication and complex conjugation as involution. This is a commutative C\(^*\)-algebra.

A \emph{classical spin model} for our Bell--CHSH configuration consists of:
\begin{itemize}
  \item \(\{\pm1\}\)-valued random variables \(A_x(\lambda)\), \(B_y(\lambda)\),
    \begin{equation}
      A_x(\lambda), B_y(\lambda) \in \{\pm1\}, \qquad x,y\in\{0,1\},
    \end{equation}
    interpreted as outcomes for directions \(a_x\) and \(b_y\);
  \item locality (factorization): \(A_x\) depends only on \(a_x\) and \(\lambda\), and \(B_y\) only on \(b_y\) and \(\lambda\); at the probabilistic level \(P(A,B|a_x,b_y,\lambda) = P(A|a_x,\lambda)\,P(B|b_y,\lambda)\);
  \item correlations given by
    \begin{equation}
      E_{\mathrm{cl}}(a_x,b_y)
      := \int_{\Lambda} A_x(\lambda)\,B_y(\lambda)\,\rho(d\lambda).
    \end{equation}
\end{itemize}
All \(A_x\) and \(B_y\) belong to the same commutative algebra \(\mathcal{C}(\Lambda)\), and the CHSH combination
\begin{equation}
  S_{\mathrm{cl}}(\lambda)
  := A_0(\lambda) B_0(\lambda)
   + A_0(\lambda) B_1(\lambda)
   + A_1(\lambda) B_0(\lambda)
   - A_1(\lambda) B_1(\lambda)
\end{equation}
satisfies \(S_{\mathrm{cl}}(\lambda)=\pm2\), hence
\begin{equation}
  |\langle S\rangle_{\mathrm{cl}}|
  := \left|\int_{\Lambda} S_{\mathrm{cl}}(\lambda)\,\rho(d\lambda)\right| \le 2.
\end{equation}

\subsection{Main non-embeddability theorem}

We can now state the main algebraic result in a compact form.

\begin{theorem}[Spinor vs.\ classical Kolmogorov algebra]
\label{thm:non_embeddability}
Let \((\mathcal{A}_{\mathrm{spin}},\omega_{\mathrm{spin}})\) be the spinor C\(^*\)-algebra and singlet state defined above for a fixed choice of Bell--CHSH directions \(a_0,a_1,b_0,b_1\). Let \((\mathcal{C}(\Lambda),\rho)\) be any commutative C\(^*\)-algebra of classical observables arising from a Kolmogorov probability space \((\Lambda,\rho)\). There exists no unital \(^*\)-homomorphism
\begin{equation}
  \Phi : \mathcal{A}_{\mathrm{spin}} \longrightarrow \mathcal{C}(\Lambda)
\end{equation}
such that:
\begin{enumerate}
  \item \emph{Spectrum preservation}: For each \(x,y\in\{0,1\}\), \(\Phi(\Sigma_A(a_x))\) and \(\Phi(\Sigma_B(b_y))\) are \(\{\pm1\}\)-valued functions in \(\mathcal{C}(\Lambda)\);
  \item \emph{Correlation preservation}: For all \(x,y\in\{0,1\}\),
    \begin{equation}
      \int_{\Lambda} \Phi\bigl(\Sigma_A(a_x)\Sigma_B(b_y)\bigr)(\lambda)\,\rho(d\lambda)
      = \omega_{\mathrm{spin}}\bigl(\Sigma_A(a_x)\Sigma_B(b_y)\bigr).
    \end{equation}
\end{enumerate}
In particular, there is no such \(\Phi\) with \(\Phi(S_{\mathrm{QIA}})\) reproducing \(\omega_{\mathrm{spin}}(S_{\mathrm{QIA}})=2\sqrt{2}\).
\end{theorem}

\paragraph{Sketch of proof.}
Under the assumptions above, the elements
\begin{equation}
  \mathsf{A}_x := \Phi(\Sigma_A(a_x)), \qquad
  \mathsf{B}_y := \Phi(\Sigma_B(b_y)),
\end{equation}
are \(\{\pm1\}\)-valued functions on \(\Lambda\). The image of the CHSH operator is the classical random variable
\begin{equation}
  S_{\mathrm{cl}}(\lambda)
  := \mathsf{A}_0(\lambda)\mathsf{B}_0(\lambda)
   + \mathsf{A}_0(\lambda)\mathsf{B}_1(\lambda)
   + \mathsf{A}_1(\lambda)\mathsf{B}_0(\lambda)
   - \mathsf{A}_1(\lambda)\mathsf{B}_1(\lambda),
\end{equation}
which satisfies \(S_{\mathrm{cl}}(\lambda)=\pm2\), hence
\begin{equation}
  \left|\int_{\Lambda} S_{\mathrm{cl}}(\lambda)\,\rho(d\lambda)\right|
  \le 2.
\end{equation}
On the other hand, correlation preservation with the singlet state implies
\begin{equation}
  \int_{\Lambda} S_{\mathrm{cl}}(\lambda)\,\rho(d\lambda)
  = \omega_{\mathrm{spin}}(S_{\mathrm{QIA}}) = 2\sqrt{2},
\end{equation}
for suitable choices of \(a_0,a_1,b_0,b_1\). This is impossible. Therefore no such \(\Phi\) exists.

\medskip
Theorem~\ref{thm:non_embeddability} is, in C\(^*\)-algebraic language, nothing but a reformulation of the Bell--CHSH theorem in the precise form given by Fine \cite{Fine1982}: the existence of a single joint probability distribution for all four outcomes is equivalent to satisfaction of all Bell--CHSH inequalities, and the singlet correlations lie outside this classical correlation polytope. Our formulation trades the language of joint distributions for that of homomorphisms between noncommutative and commutative algebras.

\section{Spin as Bivector in Clifford and QIA Representation}
\label{sec:spinor_algebra}

In this section we introduce the minimal amount of Clifford and QIA structure needed to support the spinor algebra used above. The key point is that the generators of spin-\(\tfrac{1}{2}\) live in the \emph{bivector} sector of the three-dimensional Clifford algebra \(\mathrm{Cl}_3\), rather than in the vector space of classical \(3\)-vectors.

\subsection{Clifford algebra \texorpdfstring{$\mathrm{Cl}_3$}{Cl\_3} and bivectors}

Let \((e_1,e_2,e_3)\) be an orthonormal basis of \(\mathbb{R}^3\). The real Clifford algebra
\(\mathrm{Cl}_3\) is generated by \(e_1,e_2,e_3\) subject to
\begin{equation}
  e_i e_j + e_j e_i = 2 \delta_{ij}, \qquad i,j \in \{1,2,3\}.
\end{equation}
As a vector space, \(\mathrm{Cl}_3\) decomposes into homogeneous grades:
\begin{itemize}
  \item grade~0: scalars,
  \item grade~1: vectors (linear combinations of \(e_i\)),
  \item grade~2: bivectors (products \(e_i e_j\) for \(i\neq j\)),
  \item grade~3: the pseudoscalar \(e_1 e_2 e_3\).
\end{itemize}
The three-dimensional subspace of bivectors (grade~2) consists of oriented planes.
A convenient basis is
\begin{equation}
  J_1 := e_2 e_3, \qquad
  J_2 := e_3 e_1, \qquad
  J_3 := e_1 e_2.
\end{equation}
These obey
\begin{equation}
  J_k^2 = -1, \qquad
  J_i J_j = - J_j J_i \ (i\neq j), \qquad
  J_1 J_2 = J_3,\quad J_2 J_3 = J_1,\quad J_3 J_1 = J_2,
\end{equation}
with the standard cyclic relations. Thus the span of \(\{J_1,J_2,J_3\}\) is isomorphic
to the quaternionic Lie algebra underlying \(\mathrm{SU}(2)\).

Given a unit vector \(n = (n_1,n_2,n_3) \in S^2\), we define the bivector
\begin{equation}
  J(n) := n_1 J_1 + n_2 J_2 + n_3 J_3.
\end{equation}
Geometrically, \(J(n)\) is the oriented plane orthogonal to \(n\), and algebraically it is the
generator of rotations around the axis \(n\).

\subsection{Spin-\texorpdfstring{$\tfrac{1}{2}$}{1/2} as a spinor module and matrix representation}

A spin-\(\tfrac{1}{2}\) degree of freedom is described by a two-dimensional complex spinor space
\(S \cong \mathbb{C}^2\) that carries a representation of the even subalgebra of \(\mathrm{Cl}_3\).
More concretely, there exists a representation
\begin{equation}
  \rho : \mathrm{Cl}_3 \longrightarrow M_2(\mathbb{C})
\end{equation}
such that the bivectors \(J_k\) are mapped to anti-Hermitian matrices \(\rho(J_k)\) with
\(\rho(J_k)^2 = -I\). A convenient choice is
\begin{equation}
  \rho(J_1) = -i\sigma_x,\qquad
  \rho(J_2) = -i\sigma_y,\qquad
  \rho(J_3) = -i\sigma_z,
\end{equation}
where \((\sigma_x,\sigma_y,\sigma_z)\) is a fixed Pauli triple. The Hermitian spin operators
along directions \(n\in S^2\) are then defined by
\begin{equation}
  \Sigma(n) := -i\,\rho(J(n))
             = n_1 \sigma_x + n_2 \sigma_y + n_3 \sigma_z.
\end{equation}
By construction,
\begin{equation}
  \Sigma(n)^2 = I, \qquad
  \Sigma(n)\Sigma(m) + \Sigma(m)\Sigma(n) = 2 (n\cdot m)\,I,
\end{equation}
and the spectrum of \(\Sigma(n)\) is \(\{\pm1\}\). Thus \(\Sigma(n)\) coincides with the spin
operator along direction \(n\) used in Sec.~\ref{sec:main_theorem}. In particular, spin
components are not classical vectors in \(\mathbb{R}^3\), but matrix images of bivectors
\(J(n)\) acting on spinors.

\subsection{Two spins, locality, and the singlet in bivector form}

For two spin-\(\tfrac{1}{2}\) particles, the total Hilbert space is
\begin{equation}
  H = S_A \otimes S_B.
\end{equation}
The bivector generators act locally on each factor:
\begin{equation}
  J_A(n) := J(n) \otimes 1, \qquad
  J_B(n) := 1 \otimes J(n),
\end{equation}
and under \(\rho\otimes\rho\) we recover the local spin operators
\begin{equation}
  \Sigma_A(n) = \Sigma(n)\otimes I, \qquad
  \Sigma_B(n) = I\otimes \Sigma(n),
\end{equation}
used in the Bell--CHSH analysis.

The singlet state \(|\Psi\rangle \in H\) can be characterized purely in terms of bivectors:
it is the unique (up to phase) spinor satisfying
\begin{equation}
  \bigl(J_A(n) + J_B(n)\bigr) |\Psi\rangle = 0
  \qquad \text{for all } n\in S^2.
\end{equation}
Equivalently, in matrix form,
\begin{equation}
  \bigl(\Sigma_A(n) + \Sigma_B(n)\bigr) |\Psi\rangle = 0
  \qquad \forall n\in S^2,
\end{equation}
which expresses the vanishing of total spin along every direction.

The standard correlation
\begin{equation}
  E_{\mathrm{spin}}(a,b)
  = \langle\Psi, \Sigma_A(a)\Sigma_B(b)\Psi\rangle
  = -\,a\cdot b,
\end{equation}
and the Tsirelson value \(2\sqrt{2}\) for the CHSH operator \(S_{\mathrm{QIA}}\) then follow
from this bivector/spinor structure, without any reference to classical spin vectors.
The noncommutative algebra generated by the \(J_k\) (or \(\Sigma_k\)) is precisely the
spinor/Clifford algebra \(\mathcal{A}_{\mathrm{spin}}\) appearing in Theorem~\ref{thm:non_embeddability}.

\subsection{QIA and VC-block realization}

In the Quantum Index Algebra (QIA) framework, the same spin structure is implemented
in a combinatorial way. The single-particle spinor space \(S \cong \mathbb{C}^2\) is realized
as the complexification of an index code space, and the matrices \(\Sigma(n)\) are expressed
as linear combinations of elementary \(2\times2\) VC-blocks \(B_1,B_2,B_3,B_4\) acting on that
index space (see \cite{QIA_ref} for explicit formulas). For two particles, the local operators \(\Sigma_A(n)\) and \(\Sigma_B(n)\) become
tensor products of such VC-block combinations acting on disjoint index slots, making
locality explicit at the QIA level.

From the point of view of the present work, the crucial fact is that QIA reproduces
exactly the same noncommutative spinor/Clifford algebra generated by the bivectors
\(J_k\), and hence the same correlation function \(E_{\mathrm{spin}}(a,b)=-a\cdot b\) and CHSH value
\(2\sqrt{2}\). The Bell--CHSH non-embeddability result of Sec.~\ref{sec:main_theorem} is
therefore a statement about the impossibility of replacing this QIA/Clifford bivector
spin algebra by any commutative Kolmogorov algebra based on a single probability
space that preserves spectra and singlet correlations.

\section{Kolmogorov Probability and Contextuality}
\label{sec:kolmogorov}

In the standard CHSH derivation, the role of probability theory is usually
left implicit. Here we make explicit which part of Kolmogorov probability is
used, and why this assumption is too strong for spin-\(\tfrac{1}{2}\) in the singlet state.

\subsection{Kolmogorov model and joint distributions}

A Kolmogorov probability model consists of
\begin{itemize}
  \item a sample space \(\Lambda\) of hidden states \(\lambda\),
  \item a \(\sigma\)-algebra of measurable subsets of \(\Lambda\),
  \item a probability measure \(\rho\) on \(\Lambda\).
\end{itemize}
A random variable is a measurable function \(X : \Lambda \to \mathbb{R}\) (or to a
finite set such as \(\{\pm1\}\)). The key structural feature is:

\begin{quote}
  If several random variables are defined on the same sample space
  \(\Lambda\), then they automatically admit a joint distribution.
\end{quote}

Concretely, if \(A_0, A_1, B_0, B_1 : \Lambda \to \{\pm1\}\) are four random variables on
the same \((\Lambda,\rho)\), then there exists a joint random vector
\[
  (A_0, A_1, B_0, B_1) : \Lambda \to \{\pm1\}^4,
\]
and one can speak of probabilities like
\[
  \mathbb{P}\bigl(A_0 = +1, A_1 = -1, B_0 = +1, B_1 = -1\bigr),
\]
whether or not such a combination of outcomes is ever observed in a single
experimental run.

This implicit existence of a joint distribution is the nontrivial content of
assuming ``one Kolmogorov space'' for all observables under consideration. Fine's theorem \cite{Fine1982} shows that, in the Bell--CHSH setup, the existence of such a single joint distribution is equivalent to the satisfaction of all CHSH inequalities.

\subsection{How CHSH uses the Kolmogorov structure}

In the CHSH setting we introduce, for four measurement settings
\(a_0, a_1\) (Alice) and \(b_0, b_1\) (Bob),
\[
  A_x(\lambda) \in \{\pm1\}, \qquad B_y(\lambda) \in \{\pm1\},
  \qquad x,y \in \{0,1\},
\]
as hidden-variable response functions. The standard Bell assumptions identify all
these as random variables on a single \((\Lambda,\rho)\):
\[
  A_0, A_1, B_0, B_1 : \Lambda \to \{\pm1\}.
\]

The CHSH expression \(S_{\mathrm{cl}}(\lambda)\) is then a function \(\Lambda \to \mathbb{R}\), and the inequality
\begin{equation}
  |S_{\mathrm{cl}}(\lambda)| \le 2 \qquad \forall \lambda \in \Lambda
\end{equation}
is simply a pointwise algebraic fact about four \(\{\pm1\}\)-valued random variables
with a joint distribution. Averaging over \(\rho\) yields
\[
  \bigl|\langle S\rangle_{\mathrm{cl}}\bigr| \le 2.
\]

Thus the CHSH inequality is not just a statement about ``probabilities''
in a vague sense; it relies crucially on the existence of a single Kolmogorov
space \((\Lambda,\rho)\) supporting all four variables \(A_0, A_1, B_0, B_1\) simultaneously.

\subsection{Incompatible spin components and the global Kolmogorov assumption}

In the spin-\(\tfrac{1}{2}\) experiment, the four variables \(A_0, A_1, B_0, B_1\) correspond to
outcomes of incompatible measurements:
\begin{itemize}
  \item \(A_0\) and \(A_1\) refer to spin components of Alice's particle along different
        directions \(a_0\) and \(a_1\), represented in the quantum/QIA theory by non-
        commuting operators,
  \item similarly for \(B_0\) and \(B_1\) on Bob's side.
\end{itemize}
Operationally, in a given run only one of \(A_0\) or \(A_1\) is actually measured, and
only one of \(B_0\) or \(B_1\).

By placing all four variables on the same \((\Lambda,\rho)\), the Kolmogorov model
silently assumes that:
\begin{quote}
  for each \(\lambda\), the values \(A_0(\lambda), A_1(\lambda), B_0(\lambda), B_1(\lambda)\)
  all exist simultaneously, even though no experiment can reveal them all at once.
\end{quote}
Equivalently, it assumes that there is a joint probability distribution for the
quadruple \((A_0, A_1, B_0, B_1)\).

However, for noncommuting spin components in the singlet state, such a
global joint distribution is incompatible with the observed (and spinor/QIA)
correlations. This can be made precise in various ways, but at a structural level the reason is that incompatible spin observables cannot be assigned context-independent pre-existing values in a single
classical probability space.

\subsection{Commutative vs.\ noncommutative algebras}

Mathematically, a Kolmogorov probability space \((\Lambda,\rho)\) with bounded random
variables is equivalent to a commutative C\(^*\)-algebra of functions \(L^\infty(\Lambda,\rho)\).
In this language:
\begin{itemize}
  \item the classical Bell model assumes that all \(A_x\) and \(B_y\) belong to one
        commutative algebra \(\mathcal{C}(\Lambda) \cong L^\infty(\Lambda,\rho)\),
  \item the spinor/QIA description tells us that the actual spin observables
        \(\Sigma_A(a_x)\) and \(\Sigma_B(b_y)\) generate a noncommutative spinor/Clifford
        algebra \(\mathcal{A}_{\mathrm{spin}}\).
\end{itemize}

The Bell--CHSH contradiction can thus be restated as a non-embeddability
result: there is no way to embed the noncommutative spin algebra \(\mathcal{A}_{\mathrm{spin}}\) and
its singlet state into a single commutative algebra \(\mathcal{C}(\Lambda)\) such that
\begin{itemize}
  \item each spin component \(\Sigma_A(a_x), \Sigma_B(b_y)\) is represented by a \(\{\pm1\}\)-valued
        random variable in \(\mathcal{C}(\Lambda)\),
  \item and the relevant singlet correlations \(E_{\mathrm{spin}}(a_x,b_y)\) are reproduced.
\end{itemize}
If such an embedding existed, the CHSH operator would necessarily satisfy
\[
  \bigl|\langle S\rangle\bigr| \le 2,
\]
in contradiction with the spinor/QIA value \(2\sqrt{2}\). This is exactly the obstruction analyzed in more abstract terms by Pitowsky's convex polytopes \cite{Pitowsky1989} and by Abramsky and Brandenburger's sheaf-theoretic treatment of nonlocality and contextuality \cite{AbramskyBrandenburger2011}.

\subsection{Contextual probability vs.\ a single Kolmogorov space}

The point is not that probability theory is ``wrong'', but that the specific
application of a single global Kolmogorov space is too restrictive for spin-\(\tfrac{1}{2}\)
in entangled states.

In spinor/QIA terms:
\begin{itemize}
  \item For each measurement context (e.g.\ a fixed pair of settings \((a_x,b_y)\)), the
        corresponding commuting observables generate a commutative subalgebra, and within that context one can speak of ordinary (Kolmogorov)
        probabilities.
  \item What fails is the attempt to glue all such contexts together into a single
        commutative algebra \(\mathcal{C}(\Lambda)\) carrying joint values for all \(A_0, A_1, B_0, B_1\)
        in a way that remains compatible with the spinor structure and
        singlet correlations.
\end{itemize}

In this sense, the ``Kolmogorov mistake'' in the usual Bell argument is
precisely the assumption that all spin components for all settings can be
represented as commuting random variables on one global classical probability
space. The spinor/QIA description makes this obstruction explicit: the true
spin algebra is noncommutative and lives in \(\mathcal{A}_{\mathrm{spin}}\), while a single global
Kolmogorov model would require embedding it into a commutative algebra
\(\mathcal{C}(\Lambda)\), which is impossible if the singlet correlations are to be preserved.

\section{Conclusion}
\label{sec:conclusion}

We have reformulated the Bell--CHSH scenario for spin-\(\tfrac{1}{2}\) entirely within the
spinor/Clifford and QIA frameworks and separated two logically distinct ingredients of the
usual ``Bell paradox'':
\begin{enumerate}
  \item the \emph{true} algebraic structure of spin-\(\tfrac{1}{2}\), encoded in the
        noncommutative spinor algebra \(\mathcal{A}_{\mathrm{spin}}\), and
  \item an \emph{extra} classical assumption that all spin components for all
        settings are jointly defined as \(\{\pm1\}\)-valued random variables on a single
        Kolmogorov probability space.
\end{enumerate}

On the spinor/QIA side, we showed that:
\begin{itemize}
  \item spin-\(\tfrac{1}{2}\) is naturally described as a spinor object, with spin operators
        \(\Sigma(n)\) realized as matrix images of bivectors \(J(n)\) in a Clifford algebra;
  \item the singlet state \(|\Psi\rangle\) in this spinor space yields the standard
        correlation \(E_{\mathrm{spin}}(a,b) = -a\cdot b\) and achieves Tsirelson's
        bound \(2\sqrt{2}\) for the CHSH operator;
  \item locality is represented as strict tensor-factor separation in
        \(\mathcal{A}_{\mathrm{spin}}\), and the QIA representation keeps all Alice/Bob
        operations local on disjoint index slots.
\end{itemize}

On the classical side, we made explicit that:
\begin{itemize}
  \item modeling all outcomes as random variables \(A_0,A_1,B_0,B_1\) on one
        Kolmogorov space \((\Lambda,\rho)\) is equivalent to placing them in a single
        commutative algebra \(\mathcal{C}(\Lambda)\),
  \item this automatically endows them with a joint distribution and enforces
        the CHSH bound \(|S|\le 2\),
  \item such an embedding of \(\mathcal{A}_{\mathrm{spin}}\) and its singlet state into
        \(\mathcal{C}(\Lambda)\) cannot preserve the spinor/QIA
        correlations, because the noncommutative spinor relations are incompatible
        with a global commutative algebra of classical random variables.
\end{itemize}

The core message can be summarized as follows:
\begin{quote}
  The Bell--CHSH violation in the singlet experiment can be understood as the failure
  of embedding the noncommutative spinor algebra of spin-\(\tfrac{1}{2}\) and its singlet state
  into any single commutative Kolmogorov algebra that assigns joint \(\{\pm1\}\)-valued
  outcomes to all spin components. Spin-\(\tfrac{1}{2}\) is fundamentally a spinor object,
  and any attempt to represent all spin components as commuting classical random
  variables on one probability space is mathematically incompatible with the
  observed correlations.
\end{quote}

In this sense, the ``paradox'' is clarified as a clean algebraic non-embeddability
statement
\[
  \mathcal{A}_{\mathrm{spin}} \not\hookrightarrow \mathcal{C}(\Lambda),
\]
with spinor/Clifford and QIA representations providing natural mathematical frameworks
in which the structure of spin and the origin of the Bell inequality violation become
particularly transparent. Issues such as which assumption (local causality, outcome definiteness, measurement independence) one should ultimately abandon remain interpretational, but the algebraic content of the contradiction is fixed by the spinor structure of spin-\(\tfrac{1}{2}\).

\bibliographystyle{spmpsci}

\end{document}